# Drug-target affinity prediction method based on consistent expression of heterogeneous data

Boyuan Liu*
School of Computer Science, Georgia Institute of Technology, Atlanta, USA
* Corresponding author: bryanliu@gatech.edu

*Abstract* - **The first step in drug discovery is finding drug molecule moieties with medicinal activity against specific targets. Therefore, it is crucial to investigate the interaction between drug-target proteins and small chemical molecules. However, traditional experimental methods for discovering potential small drug molecules are labor-intensive and time-consuming. There is currently a lot of interest in building computational models to screen small drug molecules using drug molecule-related databases. In this paper, we propose a method for predicting drug-target binding affinity using deep learning models. This method uses a modified GRU and GNN to extract features from the drug-target protein sequences and the drug molecule map, respectively, to obtain their feature vectors. The combined vectors are used as vector representations of drug-target molecule pairs and then fed into a fully connected network to predict drug-target binding affinity. This proposed model demonstrates its accuracy and effectiveness in predicting drug-target binding affinity on the DAVIS and KIBA datasets.**

## I. Introduction

Drug discovery is the process of discovering potential new drugs. The process of drug discovery usually requires huge human and material investment and time consumption and involves knowledge of biology, pharmacy, chemistry, and other fields. In the whole process of drug development, related research on drug-target interactions plays an important role throughout. Drug targets are the main targets of drug molecules and biological molecules that have a causal relationship with the occurrence of diseases in the body, usually biological macromolecules such as proteins and nucleic acids. Studying drug-target interactions is of great significance in the process of virtual screening, drug lead discovery, drug repositioning, and drug activity studies.

Exploring drug-target interactions is a critical step in drug repositioning. Hopkins et al. [1] proposed network pharmacology, explaining that the interaction between drug molecules and drug targets is not a specific one-to-one correspondence, but a many-to-many network. The process of new uses is called drug repositioning, and the technology avoids the expensive and lengthy drug development process [2]. Exploring the interaction between drug targets and drugs is a key step in drug repositioning. Potential effective drug molecules are screened out through high-throughput drug-target binding assays on a drug target.

Exploring the interaction mechanism between drug molecules and targets can also help to optimize lead compounds [3]. Lead compounds are small molecules with specific structures and drug activities screened from many compounds, which are the initial steps in the development of new drugs. However, some leads have defects such as high toxicity, unstable chemical structure, insufficient chemical activity, and unreasonable pharmacokinetics. By studying the chemical structure of drugs with similar properties to the lead, the chemical structure of the lead is modified. The process of optimizing the structure of a lead to meet the requirements of a drug candidate is called lead optimization [4]. By studying the drug-target interaction, we can deeply understand the binding conformation, mode, affinity, and other information between the drug and the target to help us understand the interaction mechanism between the drug and the target and guide the optimization of the lead.

At present, the drug-target interaction study by experimental methods such as nuclear magnetic resonance spectroscopy [5] and surface plasmon spectroscopy [6] cannot be effectively implemented due to the problems of capital, time cost, and low throughput [7-8]. High-throughput screening experiments are expensive and time-consuming [9-10]. And an exhaustive search strategy is not feasible because there are millions of compounds that are structurally similar to small drug molecules [11] and tens of thousands of potential drug targets [12-13]. With the continuous expansion and improvement of drug molecule databases and protein target databases, there is an urgent need to explore drug-target interactions through computational methods, and the potential is vast. Therefore, it is of practical significance and research value to establish a computational model to predict new drug-target interactions based on known drug-target experimental results.

## II. Related work

The main methods for predicting drug-target interactions include the molecular docking method, molecular dynamics simulation, machine learning method, and deep learning method. The advantages and limitations of each method are discussed in this section.

## A. Molecular docking method

Molecular docking technology [14] is a method for predicting the binding mode and binding affinity of biomolecules based on the three-dimensional structure of receptors and ligands through computer simulation of the interaction between biomolecules. As shown in Figure 1, molecular docking calculation is the process of simulating the formation of optimal conformations between molecules according to the principles of geometric structure complementarity and energy matching. Due to its strong interpretability, this technology is mainly used in the early virtual screening of potential drug molecules and in the research of applying reverse docking technology to find potential drug targets. However, molecular docking technology also faces obstacles. Obtaining 3D data on drug molecules is itself a challenging problem, and the process of large-scale molecular docking calculations is often time-consuming.

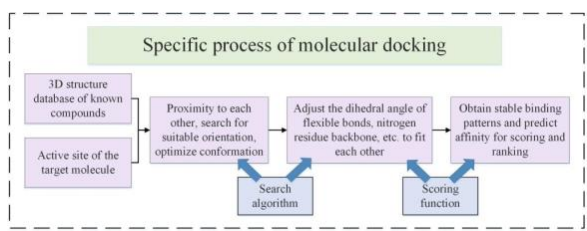

*Figure 1.* The specific process of molecular docking.

## B. Molecular dynamics simulation

Molecular dynamics simulation [15] is a method of predicting intermolecular interactions by simulating the physical motion trajectories and electronic interactions of atoms and molecules by computer. The binding free energy between ligand-receptor molecules can be calculated, and lead drugs can be screened out by analyzing the calculation results. Compared with the molecular docking method, this method can better consider the flexibility of biomolecular structure and more accurately predict the drug-target binding mode and its binding affinity [16]. Molecular dynamics simulation methods also require accurate information on drug molecules, and it is often difficult to obtain these data, and molecular dynamics simulation methods often consume more computing resources and time costs, so molecular dynamics methods also have greater limitations.

## C. Machine Learning Methods

Machine learning methods show great promise for drug-target interaction prediction because they allow large-scale testing of drug candidates in a relatively short period of time. Usually, machine learning methods do not require the intervention of much professional knowledge, and the model can automatically acquire hidden features in the data for prediction through learning. Therefore, machine learning methods have significant advantages in drug-target prediction tasks. Wang et al. [17] used a support vector machine model to predict whether there is an interaction between a drug and a target by inputting descriptors generated based on the chemical structure of the drug and protein sequence information. Li et al. [18] proposed a random forest-based model. The molecular docking method of this model applies the Kronecker similarity matrix product, using the similarity scores in KronRLS [19] as input, to improve prediction performance.

Databases related to biomolecules are constantly expanding. Traditional machine learning methods such as support vector machines and logistic regression often enter a bottleneck period in the case of large data volumes, limiting the accuracy of drug-target prediction tasks. Furthermore, the input of machine learning methods is usually molecular descriptors processed by experts in pharmacology and biology, and the inability to learn useful molecular representations from raw data features is also a barrier that limits machine learning methods.

## D. Deep Learning Methods

As a branch of machine learning methods, deep learning model architectures are more flexible and outperform traditional machine learning methods on large-scale data. The emergence of many large biomedical databases such as PubChem [20], NCI [21], ChEMBL [22], and PDB [23] has laid a solid foundation for the application of deep learning in the field of drug discovery. Deep learning methods are used to predict drug absorption, distribution, metabolism, toxicity, activity prediction, and protein-protein interaction. Good results have been achieved in the tasks of prediction of molecular and molecular mechanical properties.

Using a deep learning network to perform representation learning on drug molecules and protein molecules as distributed input, and then using a simple classifier network to perform binary classification or regression tasks is the basic framework for exploring drug-target interactions. Wang et al. [24] used stacked autoencoders for protein molecule representation learning to mine high-dimensional features of protein sequences, mapped protein molecule features into a set of vectors, and connected drug molecule fingerprints into support vector machines or random forest models for binary classification prediction. Wallach et al. [25] proposed the AtomNet deep learning model to predict the binding affinity between drug molecules and targets. AtomNet uses a convolutional neural network to perform local convolution operations on the target structure to extract the complex chemical features of the target protein. It makes full use of the structural information of the drug molecule and the target protein to accurately predict drug-target binding affinity. Tsubaki et al. [26] applied the word embedding technology commonly used in natural language processing to the feature extraction of protein sequences and drug molecular fingerprint sequences, aiming to map the protein sequence and molecular fingerprint sequence information into a set of vectors in space, so that similar proteins or drug molecules are closely distributed in the vector space, and the learned representation is used as input to predict drug-target interactions.

## III. METHODOLOGY

The overall design of the drug-target binding affinity prediction model is shown in Figure 2. In this paper, an end-to-end prediction model is proposed. Based on the SMILES sequence of the drug molecule and the amino acid sequence of the target protein as input, the characterization information of drug-target binding affinity such as dissociation constant ($K_i$), inhibition constant ($K_d$), the value of the median inhibitory concentration ($IC_{50}$) can be predicted. The structural information of the drug molecule is extracted from the SMILES sequences using the cheminformatics software RDkit to generate the drug molecule graph and adjacency matrix. The drug molecule structure graph is fed into the graph neural network model to generate the drug molecule embedding. Similarly, the GRU model is fine-tuned with target protein sequence data to generate an embedding of the target protein sequence. The drug molecule embedding and the target protein embedding are spliced into a longer vector to characterize the information of the drug-target pair, and passed to the fully-connected neural network to perform a regression task to predict the binding affinity of the drug-target pair.

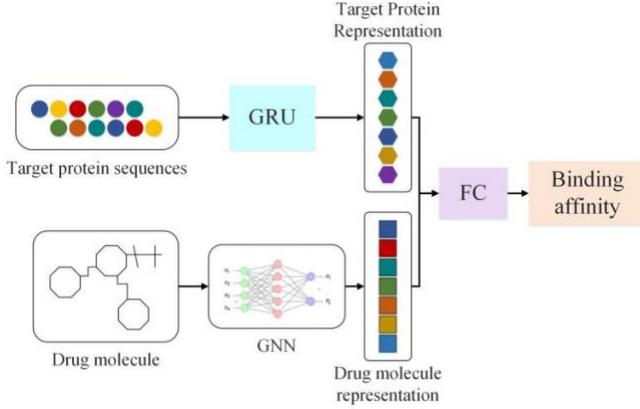

*Figure 2.* The end-to-end prediction model proposed in this paper.

### A. GRU

The structure of GRU is similar to the structure of LSTM, except that the recurrent unit of GRU has only two input variables: the input at the current moment and the state of the hidden layer at the previous moment. GRU can achieve the same functions as LSTM by introducing only two gating mechanisms [27], an update gate and a reset gate. Its internal structure is shown in Figure 3.

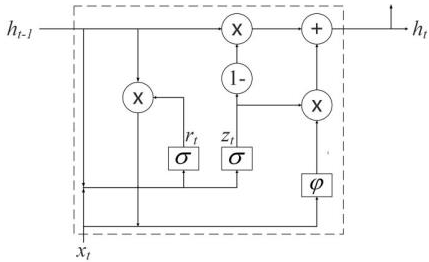

*Figure 3.* Internal structure of the GRU circulator unit.

$\tilde{h}_t$ is the activation state of the hidden layer at the current moment, $\varphi$ is the Tanh excitation function, $h_t$ is the state of the hidden layer at the current moment, $h_{t-1}$ is the state of the hidden layer at the previous moment, $r_t$ is the reset gate, and $z_t$ is the update gate. As shown in Figure 3, there are only two gating mechanisms in the recurrent unit of the GRU neural network, the reset gate $r_t$, and the update gate $z_t$. The input of the current moment and the state of the hidden layer of the previous moment are used as input information, which is multiplied with the weight matrix and processed by the sigmoid function to obtain the reset gate and the update gate. The update gate characterizes the degree of information flow from the previous moment to the current moment, and the closer its size is to 1, the greater the degree of information retention in the previous moment. The mathematical expression of the update gate calculation process is:

$$net_z = w_z x_t + u_z h_{t-1} \quad (1)$$
$$z_t = \sigma(net_z) \quad (2)$$

where $w_z$ and $u_z$ are the weights, $x_t$ is the input at the current moment, $\sigma$ is the sigmoid excitation function, and $net_z$ is the input to the excitation function. The reset gate characterizes the degree of forgetting of the information at the previous moment, and the closer its size is to 0, the greater the degree of forgetting of the information at the previous moment. The mathematical expression of the reset gate calculation process is:

$$net_r = w_r x_t + u_r h_{t-1} \quad (3)$$
$$r_t = \sigma(net_r) \quad (4)$$

where $w_r$ and $u_r$ are the weights and $net_r$ is the input of the excitation function. After getting the update gate and the output gate, the reset gate is applied to the state of the hidden layer at the previous moment, and the result is combined with the input information at the current moment. The summation process is carried out after introducing the weight matrix. The result after the summation process is passed through the Tanh excitation function to obtain the activation state of the hidden layer at the current moment, which is calculated by Equation 5 and 6:

$$net_{\tilde{h}} = w x_t + u(r_t \odot h_{t-1}) \quad (5)$$
$$\tilde{h}_t = \varphi(net_{\tilde{h}}) \quad (6)$$

where $w$ and $u$ are the weights, and $\odot$ is the Hadamard product of the matrix. The state of the hidden layer at the previous moment and the activation state of the hidden layer at the current moment are processed simultaneously by the update gate, and the results are summed to obtain the state of the hidden layer at the current moment, and the mathematical expression can be expressed as:

$$h_t = (1 - z_t) h_{t-1} + z_t \tilde{h}_t \quad (7)$$

The output of the implied layer $h_t$ is calculated by analyzing the input information, and the output of the entire neural network can be obtained through the operation of the output layer of the neural network after the output of the implied layer is obtained, and the formula of the output layer can be expressed as:

$$net_y = w_y h_t \qquad (8)$$
$$y = \sigma(net_y) \qquad (9)$$

where $w_y$ is the weight, $y$ is the 512-dimensional target protein sequence feature output by the neural network, and $net_y$ is the input to the excitation function.

### B. GNN

The GIN model proposed by Xu et al. [28] does not include edge features, so some improvements are needed to update the GIN so that it takes into account edge features as well. Similarly, the same operation is taken for GAT and GCN to incorporate edge features. First, for the node input and edge input containing two-dimensional category features, the category features of each dimension of the node features are mapped to a 300-dimensional embedding by the embedding layer. The two-dimensional embeddings are summed to eventually obtain a 300-dimensional node input. The same operation is performed for the edge features. For the iteration function of each layer of GNN, both node features and edge features are considered, as in Equation 10:

$$h_v^{(k)} = \mathrm{Re}lu(MLP^{(k)}(\sum_{u \in N(v) \cup v} h_u^{(k-1)} + \sum_{e=(v,u):u \in N(v) \cup \{v\}} h_e^{(k-1)})) \qquad (10)$$

$N(v)$ is an adjacent node of $v$, $h_u^{(k-1)}$ represents the node feature of layer $k-1$, and $h_e^{(k-1)}$ represents the edge feature of the previous layer, and it can be noticed in the equation that a self-loop is added to the node. Based on Equation 10, the node information can then be updated with the edge information.

The settings of each graph neural network model are described in this section. The number of GIN layers is chosen as 5. In the task of contextual relationship graph prediction, $r_1$ is set as 4, and $r_2$ is set as 7. A three-layer GIN is used as an auxiliary GIN to perform feature extraction on the contextual relationship graph. The number of GAT layers is 2, and the number of attention heads is chosen as 2. The number of GCN layers is 2, and the auxiliary GCN and GAT are both 2 layers. In the attribute masking task, 15% of the nodes are randomly selected and masked. They are labeled using node category features. In training, batch normalization [29] is applied before the Relu layer. The batch size is selected as 256 for the self-supervised pre-training task and 32 for the supervised pre-training task. The dropout [30] method is applied to prevent overfitting for all $h_v^{(k)}$ except for the output layer, with a dropout rate of 20%. The Adam [31] optimizer is used for parameter learning with a learning rate of 0.001 and an output dimension of 300.

## IV. EXPERIMENT

The DAVIS dataset [32] contains the results of experimental determinations of the interaction between kinase protein families and their associated inhibitors, as well as the corresponding dissociation constants $K_d$ values. Specifically, the DAVIS dataset contains interaction strength data for 442 target proteins and 68 compound ligands with $K_d$ values ranging from 5.0 to 10.8. The KIBA dataset [33] uses the KIBA score to characterize drug-target binding affinity, which is constructed by statistically analyzing the kinase inhibitor-related activity metrics $K_i$. The KIBA dataset consists of 467 target proteins and 52498 small drug molecules. The filtered KIBA dataset consists of 2111 mutually exclusive small drug molecules and 229 mutually exclusive target proteins.

The regression task uses a three-layer fully connected neural network with dimensions of 1024, 1024, and 512, respectively, using the Relu activation function. A dropout layer is added after the first two layers of the fully connected network. The experimental batch size was chosen as 256 and the dropout rate was chosen as 0.2. Since the binding affinity prediction was a regression task, the mean square error (*MSE*) was chosen as the loss function with the following loss function equation:

$$MSE = \frac{1}{n}\sum_{i=1}^{n}(P_i - Y_i)^2 \qquad (11)$$

where $P$ is the predicted vector value, and $Y$ is the actual affinity value.

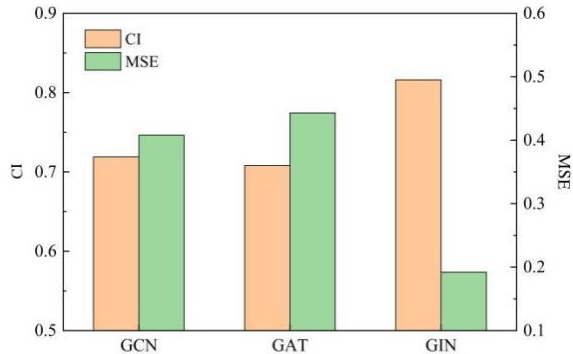

*Figure 4.* Results of the GCN, GAT and GIN models on the KIBA dataset. The GIN model achieved the highest concordance index and the lowest mean squared error, indicating better feature extraction ability on the molecular graph data than the GCN and GAT model.

The concordance index (CI) and MSE results of the GCN, GAT, and GIN models on the KIBA dataset are shown in Figure 4 above. The GIN model achieved the highest concordance index and the lowest mean squared error, indicating higher consistency between the predicted and target values. This result demonstrates the better feature extraction ability of GIN on the molecular graph data.

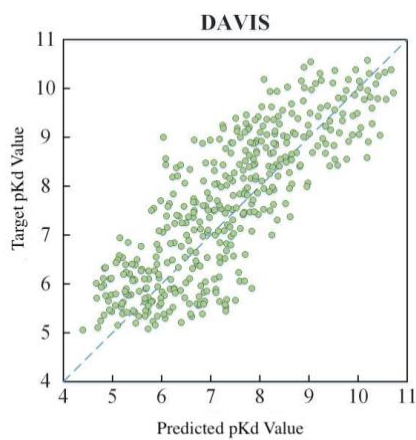

*Figure 5.* Prediction distribution of the DAVIS dataset. A large number of predictions are distributed within the interval of $pK_d$ values from 5 to 6 and the interval from 7 to 9, which is consistent with the data distribution in the DAVIS dataset.

To better visualize the correlation between the predicted and target values, the distribution of the prediction results on the DAVIS and KIBA datasets are shown in Figure 5 and 6, respectively. An ideal prediction model should keep the predicted value (p) close to the target value (m), and the distribution of the predictions should locate near the dotted line (p=m) in the figures. For the DAVIS dataset, the majority of predictions were distributed in the interval of $pK_d$ values from 5 to 6 and the interval of 7 to 9, while the rest were scattered sparsely in the area. On the other hand, the predictions on the KIBA dataset were distributed densely around the dotted line, indicating a stronger correlation between the predicted values and the target values. The model had better performance on the KIBA dataset than on the DAVIS dataset, which could be caused by model complexity and dataset size. The complex GIN model is more likely to favor the larger KIBA dataset than the DAVIS dataset. Overall, satisfactory results were achieved on both datasets, demonstrating the desirable performance of the proposed model in this paper.

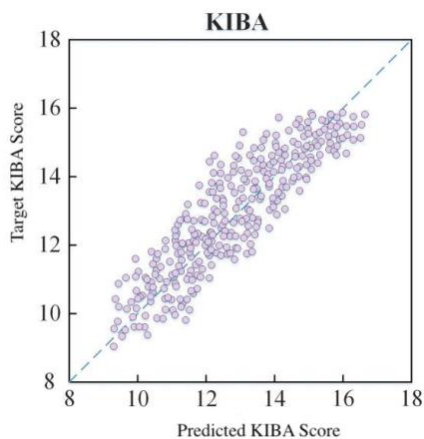

*Figure 6.* Prediction distribution of the KIBA dataset. The predictions on the KIBA dataset distributed densely around the dotted line, indicating a stronger correlation between the predicted values and the target values.

## V. CONCLUSION

Whether, how, and in what way a drug molecule binds to a target protein is an essential guide for drug discovery and analysis. In this paper, we used drug-target interactions as a starting point to investigate how to apply deep learning methods to accurately predict the binding affinity, an important indicator of drug-target interactions. An end-to-end model applying a graph neural network to extract drug molecule features and GRU to extract target protein features is proposed and validated on DAVIS and KIBA datasets. The graph neural networks demonstrated great potential in the experiments. It is worth continuing to explore their potential on the drug-target affinity prediction task in future work.